\documentstyle[multicol,prl,aps]{revtex}

\input{epsf}

\begin{document}
\draft

\title{
Theory of composite-band Wannier states and 
order-N electronic-structure calculations
}
\author{Takeo Hoshi and Takeo Fujiwara}
\address{
Department of Applied Physics, 
University of Tokyo, Bunkyo-ku,
Tokyo, Japan}

\date{\today}
\maketitle

\begin{abstract}
From the order-N electronic-structure formulation,
a Hamiltonian is derived,
of which lowest eigen state is 
the generalized or composite-band Wannier state.
This Hamiltonian maps 
the locality of the Wannier state to 
that of a virtual impurity state 
and to a perturbation from a bonding orbital.
These theories are demonstrated
in the diamond-structure solids,
where the Wannier states are constructed 
by a practical order-N algorithm with the Hamiltonian.
The results give a prototypical picture of 
the Wannier states in covalent-bonded systems.
\end{abstract}

\begin{multicols}{2}


Recently, 
Wannier state (WS) has been re-focused 
as a foundation of the order-N methods,
linear-scaling methods 
for large-scale electronic-structure calculations
\cite{ORDER-N,MAURI}.
The original concept of WS's is defined 
within a single isolated band under the periodic boundary,
but now is generalized to systems with composite bands 
and/or non-periodic boundaries.
\cite{ORDER-N,MAURI,KOHN93,MAXWANI}.
This letter is devoted to the fundamental theories
of WS's in composite-band systems.


The generalized WS's can be defined, in insulators,
as localized one-electron states that satisfy 
\begin{eqnarray}
H \psi_k = \sum_{j=1}^{N} \varepsilon_{kj} \psi_j,
 \label{SCE-UNITARY}
\end{eqnarray}
where $N$ is the number of occupied states.
Equation (\ref{SCE-UNITARY}) is derived from 
a variational procedure within a single Slater determinant.
The parameters $\varepsilon_{ij}$ are the Lagrange multipliers
for the orthogonality constraints
$ \langle \psi _i | \psi _j \rangle = \delta_{ij}$
and satisfy $\varepsilon_{ji} =  \langle \psi _i | H | \psi _j \rangle$.
The above definition does not uniquely determine the wavefunctions.
The resultant set of one-electron states $\{ \psi_i \}$
has a \lq gauge' freedom in the sense that 
any physical quantity is invariant 
under the unitary transforms with respect to the occupied states 
$ \psi_i \rightarrow \psi'_i \equiv \sum_{j=1}^{N} U_{ij} \psi_j$,
where $U$ is a unitary matrix.
If this \lq gauge' is fixed so as to diagonalize the matrix 
$\varepsilon_{ij}$,
we obtain the set of the eigen states
$\{ \psi_{k}^{\rm (eig)} \}$,
or the Bloch states in the periodic boundary;
$H \psi_k^{\rm (eig)} = 
\varepsilon_{k}^{\rm (eig)} \psi_k^{\rm (eig)}$,
where 
$\{ \varepsilon_{k}^{\rm (eig)} \}$
are the eigen energies.
With WS's, the matrix $ \varepsilon_{ij}$ is not diagonal, 
but its trace gives the correct band-structure energy
$E_0 \equiv \sum_{k=1}^{N} \varepsilon_{k} ^{\rm (eig)}
 = \sum_{k=1}^{N} \varepsilon_{kk}$.


The diamond-structure solids, 
C, Si, Ge and $\alpha$-Sn,
are typical composite-band systems.
For such materials,
nearest-neighbor tight-binding (TB) 
Hamiltonians can be constructed 
within sp$^3$-hybridized orbitals,
where the hopping along a bond is dominant.
The corresponding hopping integral is the half of 
the difference between the energy level of 
an antibonding orbital 
($\varepsilon_{\rm a}$)
and that of a bonding orbital
($\varepsilon_{\rm b}$);
$\Delta_{\rm ab} \equiv
\varepsilon_{\rm a}- \varepsilon_{\rm b}$.
If all the other hoppings are ignored,
the TB Hamiltonian is diagonal with respect to 
bonding and antibonding orbitals;
\begin{eqnarray}
H_0 = \sum_{k=1}^N
\left( | {\rm b}_k \rangle \varepsilon_{\rm b} \langle {\rm b}_k | + 
 | {\rm a}_k \rangle \varepsilon_{\rm a} \langle {\rm a}_k | \right)
 \label{H0}.
\end{eqnarray}
Here the $k$-th bonding and antibonding orbitals 
$| {\rm b}_k \rangle, | {\rm a}_k \rangle $
are defined by the pair of sp$^3$ orbitals on the $k$-th bond.
The WS's for $H_0$ are just 
the bonding orbitals $\{| {\rm b}_k \rangle  \}_{k=1,N}$,
and this simple picture is 
the starting point of the present theory.
In the present iterative calculations of the WS's,
the bonding orbitals are chosen as the initial states.
Since all the bonds are symmetrically equivalent
in the diamond structure,
the resultant energy levels of the WS's
$\{ \varepsilon_{kk} \}$
have the unique value
$ \varepsilon_{\rm WS} \equiv 
 (1/N) \sum_{j=1}^N \varepsilon_{j}^{\rm (eig)}$,
which is the weighted center of the valence band.

Another important hopping in the diamond structure is 
the hopping within an atom,
whose energy is one fourth of 
the energy difference between 
the atomic p-level ($\varepsilon_{\rm p}$)
and the s-level ($\varepsilon_{\rm s}$);
$\Delta_{\rm ps} \equiv 
\varepsilon_{\rm p}- \varepsilon_{\rm s}$.
Within the present TB parameterizations,
the electronic structures of the diamond-structure solids
can be scaled with the unique parameter,
$\alpha_{\rm m} \equiv \Delta_{\rm ps} / \Delta_{\rm ab}$,
called \lq metallicity' 
and a system would be metallic,
when $\alpha_{\rm m} \rightarrow 1$ 
\cite{HARRISON,HARRISON2}.
For classification of group IV elements,
some TB parameterizations were picked out 
and we obtained
$\alpha_{\rm m}=0.44$ for C \cite{CHADI-COHEN75}, 
$\alpha_{\rm m}=0.75$ \cite{CHADI79} for Si,
and 
$\alpha_{\rm m}=0.77,$ for Ge \cite{CHADI-COHEN75}.
In the present numerical demonstrations,
we use the nearest-neighbor TB Hamiltonians $H$ for Si
whose parameters are from Ref. \cite{KWON}.
Here $\Delta_{\rm ps}$ is fixed to be 6.45 eV.
$\Delta_{\rm ab}$
and all the other interatomic hoppings 
are functions of the bond length $d$.  
In the equilibrium case ($d=d_0 \equiv 4.44$a.u.),
$\Delta_{\rm ab} = 8.25$eV and $\alpha_{\rm m} =0.78$.


Equation (\ref{SCE-UNITARY}) is closely related to 
the localized-orbital order-N formulation \cite{MAURI},
where an energy functional 
$E_{\rm O(N)}  
= \sum_{i,j}^{N} 
(2 \delta_{ij} - \langle  \psi_j |  \psi_i  \rangle )
\langle \psi_i | \Omega | \psi_j  \rangle
={\rm Tr} [(2\rho - \rho^2) \Omega]$
is iteratively minimized. 
Here $\Omega \equiv H - \eta$, and 
$\rho \equiv \sum_{k=1}^{N} | \psi_k \rangle \langle \psi_k|$. 
$\rho $ is the one-body density matrix and
the energy parameter $\eta$ must be chosen to be sufficiently high
($\eta > \varepsilon_N$).
The WS wavefunctions satisfy
\begin{eqnarray}
0 &=&  \frac{\delta E_{\rm O(N)}}{\delta \langle \psi _k |}
   = ( 2  \Omega - \rho \Omega - \Omega \rho ) 
| \psi _k \rangle \nonumber \\
  &=&  2 \left( H - H_{\rm occ} \right) | \psi_k \rangle
 - 2 \eta ( 1 -  \rho )   | \psi _k \rangle
 \label{MGC-EQ}.
\end{eqnarray}
$ H_{\rm occ}$ is the Hamiltonian within 
the valence or occupied Hilbert space
\begin{eqnarray}
 \rho H = H \rho =  H_{\rm occ} \equiv
 \sum_{j=1}^{N} | \psi_j^{\rm (eig)} \rangle 
 \varepsilon_j^{\rm (eig)}
 \langle \psi_j^{\rm (eig)} |.
\end{eqnarray}
On the other hands,
Eq.(\ref{SCE-UNITARY}) and  the orthogonality constraint are rewritten as
$\left( H - H_{\rm occ} \right) | \psi_k  \rangle =0$
and $\left( 1 -  \rho \right)  | \psi _k \rangle = 0$, 
which satisfy Eq.(\ref{MGC-EQ}).
In the present numerical calculations,
the value $\eta=5$a.u. is chosen.


From Eq. (\ref{MGC-EQ}), 
we derive an eigen-value equation;
\begin{eqnarray}
 H_{\rm WS}^{(k)}  | \psi _k \rangle 
 = \varepsilon_{kk}  | \psi _k \rangle
 \label{IMPURITY5},
\end{eqnarray}
where 
\begin{eqnarray}
 & & H_{\rm WS}^{(k)}  \equiv
  H - \bar{\rho}_k \Omega - \Omega \bar{\rho}_k  \\ 
 & & \bar{\rho}_k \equiv 
 \rho -  | \psi _k \rangle \langle  \psi _k | =
 \sum_{j (\ne k)}^{N} 
 | \psi _j \rangle \langle  \psi _j |.
 \label{IMPURITY6}
\end{eqnarray}
This eigen-value problem corresponds to 
the variational procedure of a specified WS ($ \psi _k$),
while all the other WS's
($ \{\psi _j \}_{j \ne k}$) are fixed.
If $\bar{\rho}_k | \psi _k \rangle = 0$ is satisfied,
Eq.(\ref{IMPURITY5}) is reduced to Eq.(\ref{SCE-UNITARY}) .
A WS is not an eigen state of $H$,
but an eigen state of $H_{\rm WS}^{(k)}$.
We call the specified state ($ \psi _k $) 
as \lq central' WS.

\begin{figure}[b]
  \epsfxsize=7cm
  \centerline{\epsffile{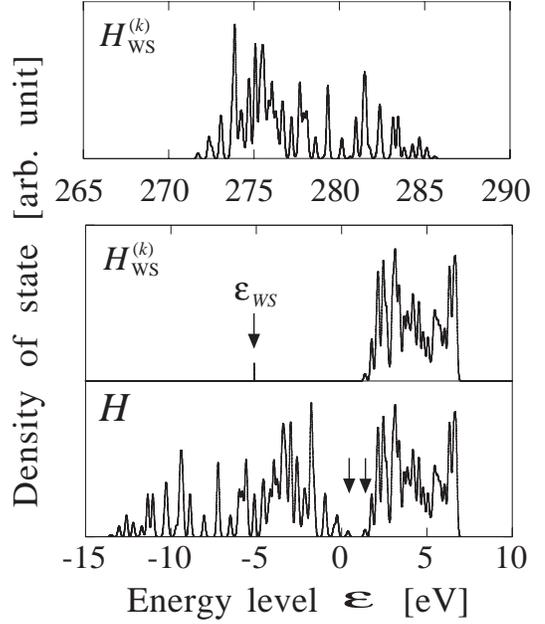}}
\caption{
The DOS
of Hamiltonians $H$ and $H_{\rm WS}^{(k)}$ for the Si case
with 512 atoms in the periodic cell.
The isolated level $\varepsilon_{\rm WS}$
is broadened by a Gaussian with a width 0.01eV and 
all the other levels are by that with a width of 0.1eV.
The two parallel arrows in $H$ indicate 
$\varepsilon_{N}^{\rm (eig)}$ and $\varepsilon_{N+1}^{\rm (eig)}$.
\label{FIG-WANI-DOS} }
\end{figure}

Figure \ref{FIG-WANI-DOS} shows 
the density of state (DOS) of $H$ and 
$H_{\rm WS}^{(k)}$ in the Si case ($d=d_0$).
We see the following properties;
(a) The \lq central' WS $ | \psi _k \rangle  $ 
is the non-degenerate ground state of $H_{\rm WS}^{(k)}$ 
with the eigen value of 
$\varepsilon_{kk} = \varepsilon_{\rm WS}$.
(b)The eigen states in the conduction band of $H$,
$( \psi _i^{({\rm eig})} ,  N+1 \le i \le 2N)$,
are also the eigen states of $H_{\rm WS}^{(k)}$
with the same eigen energies 
\begin{eqnarray}
 H_{\rm WS}^{(k)}  | \psi _i^{({\rm eig})} \rangle = 
 H | \psi _i^{({\rm eig})} \rangle
 = \varepsilon_i^{\rm (eig)} | \psi _i^{({\rm eig})} \rangle.
 \label{WS-HAMI-b}
\end{eqnarray}
(c) All the other $(N-1)$ occupied WS's of $H$ are 
{\it not} eigen states of $H_{\rm WS}^{(k)}$.
They form a high-energy band,
located at $\varepsilon \ge 2 \eta -  \varepsilon_{N} \approx 272$ eV.
The corresponding DOS profile is 
$D_{\rm high}(\varepsilon) \approx D_{\rm val}(2 \eta - \varepsilon)$,
where $D_{\rm val}(\varepsilon)$ is 
the DOS profile of the valence band of $H$.
This property rises from the relation
\begin{eqnarray}
\langle  \psi _i |  H_{\rm WS}^{(k)}  | \psi _j \rangle 
=  2 \eta - \langle  \psi _i | H_{\rm occ} | \psi _{j} \rangle,
\quad i,j \ne k.
 \label{WS-HAMI-c}
\end{eqnarray}
Due to the large energy shift of $D_{\rm high}(\varepsilon)$,
the Hilbert space of the other WS's 
is automatically 
\lq excluded' in the variational freedom 
of the WS $\psi _k$.


To construct WS's,
we used an iterative order-N algorithm 
with the Hamiltonian $H_{\rm WS}$.
The periodic cell contains 4096 atoms and
$N=8192$ doubly-occupied WS's.
For each WS $ \psi _k$,
the localization center was chosen at the center of 
the initial bonding orbital $ |{\rm b}_k \rangle$ and 
each WS was expanded into 614 sp$^3$ orbitals 
with a spherical cutoff from the localization center. 
The resultant energy per WS
($\varepsilon_{\rm WS}$) has 
a deviation of about 0.006 eV (0.1\%) from the correct value,
where the correct value is obtained 
by a standard diagonalization method 
with the primitive cell and many k-points.
The actual procedures are followings;
(i) With proper initial states of WS's, 
the density matrix $\rho$ and 
the Hamiltonians $\{H_{\rm WS}^{(k)}\}$ are constructed.
(ii) For each WS $| \psi _k \rangle$,
the one-body energy
$\langle \psi _k |H_{\rm WS}^{(k)} | \psi _k \rangle$
is minimized 
under the localization constraint on $| \psi _k \rangle$
with the fixed Hamiltonian $H_{\rm WS}^{(k)}$.
(iii) The updated WS's 
$\{\psi _k\}_k$ are orthogonalized 
using the L\"owdin symmetric orthogonalization \cite{LOWDIN}.
Then the procedure goes back to (i), untill converges.
Since the present TB Hamiltonians are upper-bounded,
WS's can be also defined for the unoccupied or conduction band.
The resultant WS's satisfy Eq.(\ref{SCE-UNITARY}),
where the $N$ one-electron states 
should be those in the unoccupied band.
Such conduction WS's can be formulated
within the energy {\it maximization} procedure of 
$E_{\rm O(N)}$ , where 
the initial WS's is chosen as 
the antibonding orbitals $\{| {\rm a}_k \rangle \}_{k=1,N}$
and the energy parameter $\eta$ 
is chosen to be enough {\it low}.

\begin{figure}[b]
  \epsfxsize=7cm
 \centerline{\epsffile{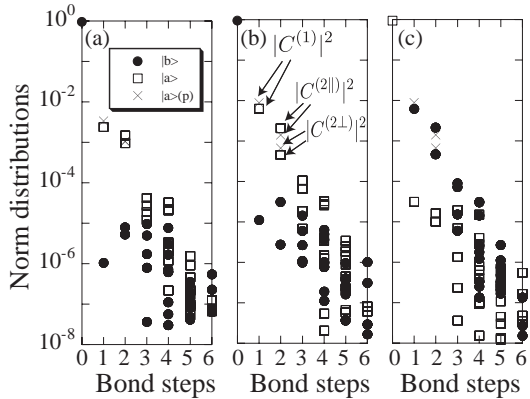}}
\caption{
Norm distributions of WS's $|C_{k \phi}|^2$, 
as a function of the bondstep from the central bond.
The closed circles and the open squares denote 
the norms on bonding and anti-bonding orbitals, respectively.
(a) the valence WS in the case with $d=0.8d_0$,
(b) the valence WS in the Si case ($d=d_0$),
(c) the conduction WS in the Si case.
The crosses denote the values from the perturbation theory.
In (a), the two values  
$|C^{(2\parallel)}|^2$ and $|C^{(2\perp)}|^2$
from the perturbation theory are almost identical.
\label{FIG-WANI-WEIGHT}
}
\end{figure}

Figure \ref{FIG-WANI-WEIGHT} shows 
the norm distributions $|C_{k \phi}|^2$ of some WS's,
where $\{\phi \} \equiv \{ {\rm b}_k, {\rm a}_k  \}$.
Figure \ref{FIG-WANI-WEIGHT} contains all the 614 
orbitals of the resultant WS's.
The case (a) is the WS with $d=0.8d_0$,
whose metallicity $\alpha_{\rm m}=0.47$
might corresponds to the Carbon case.
The case (b) is the equilibrium Si case ($d=d_0$).
For both cases,
the norm distributions on bonding orbitals 
are very small except the central one,
because they are occupied mainly by the other WS's.
The conduction WS's in the Si case
is also plotted in Fig.\ref{FIG-WANI-WEIGHT}(c),
which shows the similar decay property 
as in the valence WS (b), 
but the role of bonding and antibonding orbitals exchange with each other.
The norm of the central bond is 
about 96 \% in (a) or 94 \% in (b) and (c).
The summation of the norms upto the bondstep of $n=2$
is about 99.8 or 99.7 \% in all the cases.


The most important issue for the practical order-N calculations
is to reproduce physical quantities 
under localization constraints on WS's.
With the exact WS's $\{ \psi_k \}$, 
the physical quantity of a one-body operator $\hat{X}$ 
can be described as $\langle  \hat{X}  \rangle
\equiv \sum_{k}^{N} \langle 
\psi_k^{\rm (eig)} | \hat{X} | \psi_k^{\rm (eig)} \rangle 
= \sum_{k}^{N} \langle \psi_k | \hat{X} | \psi_k \rangle $.
In a practical order-N calculation 
with a localization constraint,
the strict orthogonality constraint is modified
to an approximate one
$( \langle  \psi_j |  \psi_i  \rangle \approx  \delta_{ij} )$
and the expression of a physical quantity is replaced by
$\langle  \hat{X}  \rangle
= \sum_{i,j}^{N} 
(2 \delta_{ij} - \langle  \psi_j |  \psi_i  \rangle )
\langle \psi_i | \hat{X} | \psi_j \rangle$,
which is mainly contributed by  
the diagonal elements $ \langle \psi_k | \hat{X} | \psi_k \rangle $.
If an operator $\hat{X}$ is short-ranged,
like the present TB Hamiltonian,
its matrix element
$\langle \psi_k | \hat{X} | \psi_k \rangle $  
should be determined dominantly by a \lq central' region of WS,
but little by the \lq tail' or 
the asymptotic long-distance behavior.
From this point of view,
we analyse the present WS's through the matrix elements
$\langle \psi_k | \hat{X} | \psi_k \rangle $ of some operators,
not through the asymptotic long-distance behavior.
Notice that the asymptotic long-distance behavior of WS 
is discussed in Refs.\cite{ORDER-N}.

If we apply the above discussion
to the operator $(\hat{\mbox{\boldmath $r$}}-
\mbox{\boldmath $r$}_k)^2$,
where $\mbox{\boldmath $r$}_k$ is the center of 
the central bond ($| {\rm b}_k \rangle$),
an effective spatial spread of WS
can be defined as
$\bar{r}_{\rm WS} \equiv
( \langle \psi_k | (\hat{\mbox{\boldmath $r$}}-
\mbox{\boldmath $r$}_k)^2 | \psi_k \rangle )^{1/2}$.
The values were actually calculated with the assumption
that each sp$^3$ orbital is localized 
at the atom that the orbital belongs to.
For a bonding orbital, this parameter is 
$\bar{r}_{\rm b} \equiv d/2$ from its definition.
For the WS in the Si case ($d=d_0$),
we obtained 
$\bar{r}_{\rm WS} = 1.16 \bar{r}_{\rm b}$.
To see the effect of the boundary condition,
we also calculated the system with 512 atoms 
{\it without} localization constraint
and found $\bar{r}_{\rm WS}= 1.19 \bar{r}_{\rm b}$.
Since the operator 
$(\hat{\mbox{\boldmath $r$}}-\mbox{\boldmath $r$}_k)^2$
is {\it not} short-ranged,
the value of $\bar{r}_{\rm WS}$ might be sensitive 
to the boundary conditions.

Another definition of the spatial spread of WS
can be derived from the Hamiltonian $H_{\rm WS}^{(k)}$.
The Hamiltonian $H_{\rm WS}^{(k)}$ 
has one localized eigen state $\psi _k$
and the conduction band,
and so $H_{\rm WS}^{(k)}$ 
maps the WS formally to an impurity state,
where  the corresponding ionization energy 
is defined as 
$\Delta_{\rm WS} \equiv
 \varepsilon_{N+1}^{\rm (eig)}- \varepsilon_{\rm WS}$.
A simplest case is the Hamiltonian in Eq.(\ref{H0}),
where the WS's are reduced to bonding orbitals
and the corresponding gap $\Delta_{\rm WS}$
is to $\Delta_{\rm ab}$.
Using the uncertainty relation,
a spatial spread is defined as
$\xi_{\rm b} \equiv \hbar/\sqrt{2m_{\rm e} \Delta_{\rm ab}}$,
where $m_{\rm e} \equiv 1$a.u.
Using the value in the Si case ($\Delta_{\rm ab}=$8.25eV),
we obtain $\xi_{\rm b}$=0.29$d_0$,
which is consistent to the fact that 
the spread of a bonding orbital
should be less than or about equal to the bond length $d_0$.
Such parameters can be also defined 
for WS's as 
$\xi_{\rm WS} \equiv 
 \hbar / \sqrt{2m_{\rm e} \Delta_{\rm WS}}  
 = \xi_{\rm b} 
 \sqrt{\Delta_{\rm ab} / \Delta_{\rm WS}} $.
For the WS in the Si case,
we obtained $\Delta_{\rm WS}=6.49$eV 
and $\xi_{\rm WS}/\xi_{\rm b} = 1.13$,
which agrees, in the order,
to the another definition 
of the spatial spread 
$\bar{r}_{\rm WS}/ \bar{r}_{\rm b}=1.16$ or $1.19$.
This agreement shows that 
the mapping theory to a virtual impurity state
is consistent to the resultant WS's.
The resultant values of the spatial spread
lead us to the conclusion 
that the WS in the Si case is so localized that 
its spatial spread is 
in the same order as that of a bonding orbital.

To see a limiting case with vanishing the bandgap,
we also calculated 
an artificial case with $\Delta_{\rm ps} =0$,
where the system is a direct-gap insulator and 
the bandgap $\Delta = 8 E_{xx}$ is located at
$\Gamma$ point \cite{TBPARA}.
We modified the parameter $E_{xx}$ from 
the value in the Si case ($E_{xx}=0.20$eV) to 
an almost vanishing one ($E_{xx}=0.0005$eV).
This modification was done by the tuning of 
$V_{pp \sigma}$ and $V_{pp \pi}$,
so as to keep $E_{xy}$ unchanged \cite{TBPARA}.
This modification changes the value of 
$\Delta_{\rm ab}$ as well.
The resultant WS still shows a localized property
with the spatial spread of 
$\bar{r}_{\rm WS}/ \bar{r}_{\rm b}=1.15$ or $1.20$,
where the latter value is from the calculation with 512 atoms 
{\it without} localization constraint.
From the gap parameters 
($\Delta_{\rm ab}=7.95$eV,$\Delta_{\rm WS}=4.23$eV),
we obtained $\xi_{\rm WS}/\xi_{\rm b}=1.37$.
These resultant values of the spatial spread 
lead us to the same conclusions as 
those in the above Si case.


The first-order perturbation of Eq.(\ref{IMPURITY5}) 
can be constructed using $H_0$ 
in Eq.(\ref{H0}) as the non-perturbative Hamiltonian; 
\begin{eqnarray}
| \psi_k \rangle = 
C^{(0)} |{\rm b}_k \rangle + 
\sum_{j (\ne k)} C^{(\nu(j))} |{\rm a}_{j} \rangle 
\label{WANI-PBS}
\end{eqnarray}
where $C^{(0)} \approx 1$.
The suffix $\nu$ specifies the bond step
and the inequivalent bond sites
from the central bond $|{\rm b}_{k} \rangle$.
In the perturbation terms,
bonding orbitals $\{|{\rm b}_{j} \rangle\}_{j \ne k} $
are \lq excluded',
because these are the other WS in the non-perturbative terms 
and are in the high-energy band in Fig.\ref{FIG-WANI-DOS}.
Because the Hamiltonian $H$ is a short-range operator,
the perturbation series in Eq.(\ref{WANI-PBS}) contain
only the 6 first-nearest-neighbor (FNN) antibonding orbitals
($|{\rm a}^{(1)} \rangle $) 
and the 18 second-nearest-neighbor (SNN) antibonding orbitals
($|{\rm a}^{(2)} \rangle $).
For the FNN antibonding orbitals,
the perturbative coefficients are given 
\cite{HARRISON} by
\begin{eqnarray}
 \frac{C^{(1)}}{C^{(0)}} \approx 
 \frac{\langle a^{(1)} | H | b_k \rangle}{-\Delta_{\rm ab}} =
 \frac{\Delta_{\rm ps}}{8 \Delta_{\rm ab}} =
 \frac{\alpha_{\rm m}}{8}
\label{WANI-COE1}
\end{eqnarray}
Here the factor $1/8$ stems from 
the {\it four} atomic coordination of the diamond structure,
which is a three dimensional effect.
On the other hand,
the SNN anti-bonding orbitals
are classified into two geometrically inequivalent bond sites;
The 6 SNN bonds are parallel to the central bond.
The other 12 SNN bonds exist, in a rough sense, 
within the plane perpendicular to the central bond.
We denote the corresponding coefficients 
$C^{(2\parallel)}$ and $C^{(2\perp)}$, respectively,
and propose the estimations of
\begin{eqnarray}
 \frac{C^{(2\lambda)}}{C^{(0)}} \approx 
 \frac{\langle a^{(2\lambda)} | H | b_k \rangle}{-\Delta_{\rm ab}}+
 \left( \frac{\alpha_{\rm m}}{8} \right)^2 ,
 \label{WANI-COE2}
\end{eqnarray}
where $(2\lambda)$ indicates $(2\parallel)$ or $(2\perp)$.
The first term is the first-order perturbation and 
its value is about
$+1/34$ for $C^{(2\parallel)}$ or $-1/27$ for $C^{(2\perp)}$.
Since this term is reduced to the ratio between 
two {\it inter-atomic } hoppings,
its value is almost unchanged 
within the diamond-structure solids
\cite{HARRISON}.
The second term in Eq.(\ref{WANI-COE2})
is responsible for the successive hopping 
of the FNN hoppings, where $C^{(0)}=1$ is assumed.
This term varies with $\alpha_{\rm m}$ and is 
essential to the distinction between
the cases in Fig.\ref{FIG-WANI-WEIGHT} (a) and (b).
Notice that, 
in the realistic cases ($\alpha_{\rm m} \le 1$),
the second term $(\alpha_{\rm m}/8)^2 \le 1/64$ 
is smaller than the first term, though not negligible.
The above coefficients
can be determined by 
Eqs.(\ref{WANI-COE1}), (\ref{WANI-COE2}) 
and the normalization condition
$(|C^{(0)}|^2 + 6|C^{(1)}|^2 + 6|C^{(2\parallel)}|^2 
+ 12|C^{(2\perp)}|^2 = 1)$.
For the conduction WS's, 
the same perturbation theory can be constructed, 
where the role of bonding and antibonding orbitals
exchange with each other. 
The resultant norm distributions
$|C^{(1)}|^2$, $|C^{(2\parallel)}|^2$ and 
$|C^{(2\perp)}|^2$
are shown as crosses in Fig.\ref{FIG-WANI-WEIGHT}.
The energy of the WS
$\varepsilon_{\rm WS} = 
\langle  \psi _k |  H  | \psi _k \rangle $
was also estimated from the perturbation results 
and the deviation from the correct value was 
about $0.06$ eV in the Si case,
which corresponds 
to 1 \% of the energy ($\varepsilon_{\rm WS}$) and 
to 10 \% of the energy difference 
from a bonding orbital
$(\varepsilon_{\rm b} - \varepsilon_{\rm WS})$.
Here we can see that the present TB Hamiltonian
is a short-range operator and 
the value of its matrix element 
$\langle  \psi _k |  H  | \psi _k \rangle $
can be well explained within a quite local area.

In conclusion,
the concept of composite-band WS's connects
the picture of \lq chemical bond' 
with the modern variational theory of electronic-structures.
In this letter,
we have shown how 
a WS is similar to and different from a bonding orbital
within the diamond-structure solids,
where the Hamiltonian $H_{\rm WS}$ plays 
a crucial role both for 
the construction and the analysis of WS's.
These theories are derived 
from the variational order-N formulation
and so are applicable to
other WS's in covalent-bonded systems and/or 
{\it ab initio} Hamiltonians.
Results of these theories 
give microscopic pictures 
for practical order-N calculations 
of large-scale systems.


This work is supported 
by a Grant-in-Aid for COE Research \lq Spin-Charge-Photon' and
by a Grant-in-Aid from the Japan Ministry
of Education, Science, Sports and Culture.
The numerical calculation was partly 
carried out by the computer facilities
at the Institute of Molecular Science at Okazaki
and at the Institute for Solid State Physics at the University of Tokyo.


%
%
%

\end{multicols}



\begin{thebibliography}{99}
\bibitem{HARRISON2}
The present definition of the metallicity is 
different from Ref. \cite{HARRISON} 
by a factor 1.11.
\bibitem{MASS}
From the general theory of impurity states,
one might think that the mass for 
$\xi_{\rm WS}$ should be an effective mass.
In general, this might be true, but 
the analysis of this letter shows that 
the WS for the diamond-structure solids
is very similar to a bonding orbital and 
such an effective mass should be that
of a bare electron (1 a.u.).
\bibitem{TBPARA}
The definitions and relations
of the present TB parameters are given in Ref. \cite{HARRISON} 
or other standard textbooks.
\end{thebibliography}
\end{document}